\documentclass[aps,pra,twocolumn,showpacs]{revtex4}
\usepackage{graphicx}
\usepackage{epsfig}
\usepackage{dcolumn}
\usepackage{bm}

\def\bra{\langle}
\def\ket{\rangle}
\def\omelam{\omega_\lambda}
\def\gamlam{\gamma_\lambda}

\begin{document}

\title{Peierls instability, periodic Bose-Einstein condensates and density waves
in quasi-one-dimensional boson-fermion mixtures of atomic gases}

\date{\today}

\author{Takahiko Miyakawa}
\affiliation{Optical Sciences Center, The University of Arizona,
Tucson, Arizona 85721, USA}
\email{tmiyakawa@arizona.optics.edu}
\author{Hiroyuki Yabu}
\author{Toru Suzuki}
\affiliation{Department of Physics, Tokyo Metropolitan University,
1-1 Minami-Ohsawa, Hachioji, Tokyo 192-0397, Japan}

\begin{abstract}
We study the quasi-one-dimensional (Q1D) spin-polarized bose-fermi mixture
of atomic gases at zero temperature.
Bosonic excitation spectra are calculated 
in random phase approximation
on the ground state with the uniform BEC, 
and the Peierls instabilities are shown to appear 
in bosonic collective excitation modes with wave-number $2k_F$ 
by the coupling between the Bogoliubov-phonon mode of bosonic atoms 
and the fermion particle-hole excitations.
The ground-state properties are calculated in the variational method, 
and, corresponding to the Peierls instability, 
the state with a periodic BEC and fermionic density waves
with the period $\pi/k_F$ 
are shown to have a lower energy than the uniform one. 
We also briefly discuss the Q1D system confined in a harmonic oscillator (HO)
potential and derive the Peierls instability condition for it.
\end{abstract}



\maketitle

\section{\label{sec:intro}Introduction}

The cooling and trapping techniques of atoms, 
used in the realization of BEC \cite{BEC,REV} 
and degenerate fermi gases \cite{DFG}, 
have recently produced quantum bose-fermi mixtures of dilute atomic gases
\cite{BFMEXP,Schreck}.
Motivated by the controllability of various experimental parameters
such as trap geometries, particle numbers, and interaction strengths
in experiments of the ultra-cold atoms,
a number of theoretical studies have been done 
to explain various experimental results
done for bose-fermi mixtures
\cite{M98,AMSTV98,TW00,MT00,MOYS00,BHS00,VPS00,MYS,RF02,
SPZM02,SMSY02,Das}.

Further experimental advances have led to the realization
of BEC in quasi-one-dimensional (Q1D) system
\cite{Schreck,Greiner01,MITLD},  
by putting atoms in cylindrical traps long enough
that the one-particle energy-level spacing in radial direction
exceeds the interatomic-interaction energy,
and the atoms can move effectively in the axial direction.
Characteristic properties of the Q1D BEC have been demonstrated
in the formation experiments of matter-wave solitons 
in repulsively/attractively interacting systems \cite{Dark,Bright}.
Also several new phenomena, which are typical in the Q1D bose system,
have been proposed theoretically: quasi-condensates with
fluctuating phases \cite{QCond} and a Tonks-Girardeau gas
of impenetrable bosons \cite{Tonks} and so on.

In electric conductors, 
the Q1D systems have been realized in systems
where electric currents can flow easier
in one specific crystal direction than in the vertical directions \cite{1DCOND}. 
One of the most fascinating phenomena in such a system is the ``Peierls instability'':
a lattice distortion 
with the period of $2k_F$ \cite{F54P55}.
This instability gives rise to a charge-density wave in the ground state, 
a collective state of electrons, which is caused by the strong correlations 
among electrons due to the phonon-electron interaction.
The origin of the effect is the divergence in phonon spectrum 
by the low-energy particle-hole (p-h) excitations near fermi surface 
with the wave-number $2k_F$, 
which give large contribution in phonon self-energy 
because of a phase-space reduction in Q1D systems.

In the present paper, 
we investigate the occurrence of the Peierls instability 
in the Q1D uniform system of atomic-gas bose-fermi mixture at $T=0$.
Different from conductors that have the periodic structures by lattices, 
it have no periodic structures originally, 
and the instability and the periodic structure in the ground state 
is brought purely by the coupling between the Bogoliubov-phonon mode
of bosonic atoms and fermion p-h excitations.

This paper is organized as follows.
In Section II, 
we introduce the model 
and derive the equations in random phase approximation 
using the Green's function method
to calculate the bosonic excitation spectra.
In Section III, 
the equations obtained in Section II are applied
for the system with the uniform ground state, 
and solved numerically.
We show the calculated bosonic excitation spectra 
and that the Peierls instability appears
in collective excitation modes 
around the wave-number $2k_F$.
The analytical estimations are also given 
for the collective excitation spectra.
In Section IV, we show that the state with periodic BEC 
and the density-wave states for both boson and fermion
(with the period of $\pi/k_F$)  
is more stable than the uniform state. 
In Section V, we briefly discuss the Peierls instability 
in the finite Q1D system confined in a HO potential \cite{MYS03}. 

\section{\label{sec:model} Model, Green's Functions and Mean-Field Basis}

We consider a system of spin-polarized atoms at $T=0$: 
$N_b$ bosons and $N_f$ fermions with masses $m_{b,f}$ each other,  
confined in an axially-symmetric HO potential 
$U_{b,f}=\frac{1}{2} m_{b,f} \left\{\omega_r^2(x^2+y^2)+\omega_a^2 z^2\right\}$ 
where $\omega_r \gg \omega_a$. 
When atoms are confined tightly enough in the $r$-direction 
that the radial ($r$) part of their wave functions 
can be approximated by that of the lowest-energy state
in the 2D HO potential 
$\frac{1}{2} m_{b,f} \omega_r^2 (x^2+y^2)$, 
then we call that the system is in the Q1D region.  
It should be realized when,
\begin{itemize}
\item The $r$-HO quanta $\hbar\omega_r$
should be higher than the boson-boson/boson-fermion interaction energies:
$\epsilon_{bb}^{3D} =4\pi \hbar^2 a_{bb} n_b^{3D} /m_b$     and
$\epsilon_{bf}^{3D} =2\pi \hbar^2 a_{bf} n_{f}^{3D} /m_r$, 
where $n_{b,f}^{3D}$ are the boson/fermion 3D-number densities and
$a_{bb,bf}$ are the boson-boson/boson-fermion s-wave scattering lengths. 
The $m_r$ in $\epsilon_{bf}^{3D}$ is a reduced mass: 
$m_r =m_{b} m_{f}/(m_{b}+m_{f})$. 
\item The $r$-HO quanta should also be higher than the fermi energy 
obtained by 
$\epsilon_F^{3D} =\hbar^2(6\pi^2n_f^{3D})^{2/3}/2m_f$ \cite{Das}.
\end{itemize}

The atomic states for the axial ($a$) degree of freedom
are described by the $z$-dependent field operators for bosons and fermions, 
$\hat{\phi}(z)$ and $\hat{\psi}(z)$;
the effective Hamiltonian for them is modeled by
\begin{eqnarray}
   && H =\int{\!dz\,} \hat{\phi}^\dagger(z) 
      \left[-\frac{\hbar^2}{2 m_b} \frac{d^2}{dz^2}
            +U_b(z) -\mu_b \right] \hat{\phi}(z) 
\nonumber\\
    &&+ \int{\!dz\,} \hat{\psi}^\dagger(z) 
      \left[-\frac{\hbar^2}{2 m_f} \frac{d^2}{dz^2} 
            +U_f(z) -\mu_f \right] \hat{\psi}(z)
\nonumber\\
    &&+ \int{\!dz\,} \hat{\phi}^\dagger(z) 
      \Bigg[ \frac{g_{bb}}{2} \hat{\phi}^\dagger(z) \hat{\phi}(z)
\nonumber\\
    &&\qquad\qquad
             +g_{bf} \hat{\psi}^\dagger(z) \hat{\psi}(z) \Bigg] \hat{\phi}(z),
\label{HBF}
\end{eqnarray}
where the $\mu_{b,f}$ are the 1D chemical potentials 
for bosons and fermions.
The effective 1D coupling constants $g_{bb}$ (boson-boson) and $g_{bf}$
(boson-fermion) in (\ref{EqA}) are related to the s-wave scattering lengths: 
$g_{bb} = 2\hbar \omega_r a_{bb}$,
$g_{bf} = 2\hbar \omega_r a_{bf}$,
which can be obtained from the 3D interaction 
by integrating out the radial parts of the wave functions \cite{Das,O98}.
It should be noted that, in the limit $\omega_a \to 0$,
we obtain the uniform system in the axial dimension.

In weak interacting systems, 
the mean field approximation should give a good starting point 
for further calculations,  
where bosons are assumed to occupy the lowest single-particle state
$\varphi_0(z)$ at $T=0$,  
and the expectation value of the boson field-operator 
(order parameter for BEC),  
$\Phi(z)=\bra \hat{\phi}(z) \ket =\sqrt{N_b}\varphi_0(z)$, 
satisfies 1D Gross-Pitaevskii equation: 
\begin{equation}
     \left[ -\frac{\hbar^2}{2m_b} \frac{d^2}{dz^2} 
            +U_{b}(z) +g_{bb}n_b +g_{bf}n_{f} \right] \Phi
     =\mu_{b} \Phi,
\label{GPeq}
\end{equation}
where the (axial) 1D boson density $n_b$ is defined by $n_b(z)=|\Phi(z)|^2$.

For fermions, the ground state is obtained by a Slater determinant 
of fermion single-particle wave functions $\psi_j(z)$
for the mean-field energies $\epsilon_j^f$ 
below the 1D fermi energy $\epsilon_F=\mu_f$,  
which are obtained from the Hartree equation:
\begin{eqnarray}
     \left[ -\frac{\hbar^2}{2m_f} \frac{d^2}{dz^2} 
            +U_f(z) +g_{bf} n_b(z) \right] \psi_{j}(z)
     =\epsilon_j^f \psi_j(z).
\label{Heqf}
\end{eqnarray}
In this state, the fermion density is obtained by 
$n_f(z)=\sum_{\epsilon_j^f \leq \epsilon_F} |\psi_j|^2$.

We are interested in the bosonic excitation modes of the system, 
which can be obtained by the bosonic Green's functions
\begin{eqnarray}
     &&\qquad\qquad i\bm{G}(x_1,x_2) =
\nonumber\\
     &&\left( 
          \begin{array}{cc}
            \langle T[ \hat{\varphi}(x_1) \hat{\varphi}^\dagger(x_2) ] \rangle &
            \langle T[ \hat{\varphi}(x_1) \hat{\varphi}(x_2)         ] \rangle \\
            \langle T[ \hat{\varphi}^\dagger(x_1) \hat{\varphi}^\dagger(x_2) ]\rangle &
            \langle T[ \hat{\varphi}^\dagger(x_1) \hat{\varphi}(x_2) ] \rangle
          \end{array} 
        \right).
\end{eqnarray}
where $x=(z,t)$ and $\hat{\varphi}(x)=\hat{\phi}(z,t)-\Phi(z)$.
They satisfy the Dyson equation: 
\begin{eqnarray}
     && \left[ i\hbar \frac{\partial}{\partial t} \bm{\tau_3} 
           -K^0_b(z_1) \bm{1} \right] \cdot \bm{G}(x_1,x_2) 
     =\hbar \delta(x_1-x_2)\bm{1}                           
\nonumber\\
     &&\qquad +\int{d^2x_3\,} \hbar\bm{\Sigma}(x_1,x_3) \cdot \bm{G}(x_3,x_2),
\label{Dysoneq}
\end{eqnarray}
where  
$K^0_b(z)=-(\hbar^2 / 2 m_b) d^2/dz^2 +U_b(z) -\mu_b$ 
is a single-particle operator
and the matrices $\bm{\tau_3}$ and $\bm{1}$ are defined by
\begin{equation}
     \bm{\tau_3} =\left( \begin{array}{cc}1&0\\0&-1\end{array} \right), \qquad
     \bm{1}      =\left( \begin{array}{cc}1&0\\0&1 \end{array} \right).
\end{equation}

\begin{figure}
\begin{center}
\includegraphics*[width=8cm,height=4.5cm]{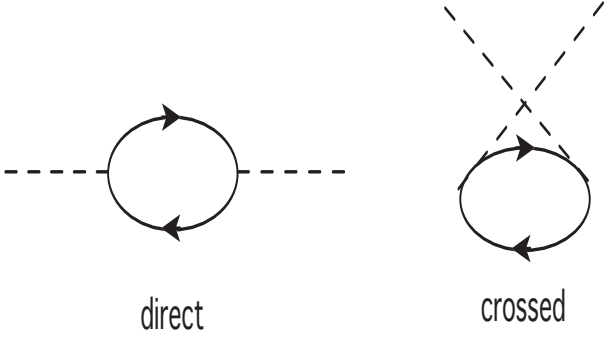}
\vspace{.3 cm}
\caption{\label{fig:1} Polarization potential induced by fermion
particle-hole excitation. Dashed line: boson (non-condensate),
solid line: fermion.}
\label{phpair}
\end{center}
\end{figure}

The boson self-energy in Eq.~(\ref{Dysoneq}) is in the $2 \times 2$-matrix form, 
and separated into the static and the dynamical parts: 
$\bm{\Sigma}(x_1,x_2) =\bm{\Sigma}_0(x_1,x_2) +\bm{\Pi}(x_1,x_2)$.
In the mean-field approximation,
the static part is given by
\begin{equation}
     \hbar\bm{\Sigma}_0 =g_{bb} 
          \left( 
          \begin{array}{cc}
               2n_{b} +\frac{g_{bf}}{g_{bb}} n_{f}&\Phi\Phi\\
               \Phi^* \Phi^*&2n_{b} +\frac{g_{bf}}{g_{bb}} n_{f}
          \end{array} 
          \right) \delta(x_1-x_2),
\end{equation}
In the dynamical part, 
we take the contributions from the fermion particle-hole (p-h) pair-excitations 
(Figure.~\ref{phpair}): 
\begin{eqnarray}
     \hbar\bm{\Pi}(x_1,x_2) 
          &=& -\frac{i}{\hbar} g_{bf}^2 
            \left( \begin{array}{cc}
                   \Phi(z_1) \Phi^{*}(z_2)&\Phi(z_1) \Phi(z_2)\\
                   \Phi^*(z_1) \Phi^{*}(z_2)&\Phi^*(z_1) \Phi(z_2)
                   \end{array} 
            \right)     
\nonumber\\
           && \times G^f(x_1,x_2) G^f(x_2,x_1),
\label{EqB}
\end{eqnarray}
which is just a polarization potential for pair-excitations.
It should be noted that this approximation is equivalent 
to the random phase approximation, 
and gives the strong correlation effects, 
especially in the bosonic collective modes 
(Bogoliubov-phonon mode).

For the fermion Green's function in (\ref{EqB}), 
we use that for the unperturbed fermion single-particle state 
obtained from Eq.~(\ref{Heqf}): 
\begin{eqnarray}
     && iG^f(x_1,x_2)
       =\sum_j \psi_j(z_1) \psi_j(z_2)^*
          e^{ -i \epsilon^f_j (t_1 -t_2) }
\nonumber\\
     &&\times [ \theta(t_1-t_2) \theta(\epsilon^f_j-\epsilon_F)
               -\theta(t_2-t_1) \theta(\epsilon_F-\epsilon^f_j)] .
\end{eqnarray}

To advance the calculation, 
we expand the boson field operators, $\hat{\varphi}$ and $\hat{\varphi}^\dagger$, 
by boson creation/annihilation operators, 
$\hat{\beta}_\lambda$ and $\hat{\beta}^\dagger_\lambda$ 
for the quasi-particle states:
\begin{eqnarray*}
     \hat{\varphi}(z) 
          =\sum_{\lambda} \left\{ u_\lambda(z) \hat{\beta}_\lambda
                           -v_\lambda^*(z) \hat{\beta}^\dagger_\lambda 
                     \right\}, \\
     \hat{\varphi}^\dagger(z) 
          =\sum_{\lambda} \left\{ u_\lambda^*(z) \hat{\beta}^\dagger_\lambda 
                           -v_\lambda(z) \hat{\beta}_\lambda 
                     \right\},
\end{eqnarray*}
where the quasi-particle wave functions, 
$u_\lambda(z_1)$ and $v_\lambda(z_1)$, 
for the eigenenergies $\hbar \Omega_{\lambda}$
can be determined from the Bogoliubov-type eigenequations:
\begin{widetext}
\begin{eqnarray}
     \mathcal{L} u_\lambda(z_1) +\int{dz_2\,}
     \left\{ n_b g_{bb} \delta(z_1-z_2)
            +\hbar \Pi(z_1,z_2;\Omega_\lambda) \right\}
          \{ u_\lambda(z_2) -v_\lambda(z_2) \}
     =\hbar \Omega_\lambda u_\lambda(z_1),     
\label{Bog1}\\
     \mathcal{L} v_\lambda(z_1) +\int{dz_2\,}
     \left\{ n_b g_{bb} \delta(z_1-z_2)
            +\hbar \Pi(z_1,z_2;\Omega_{\lambda}) \right\}
          \{ v_\lambda(z_2) -u_\lambda(z_2) \}
     =-\hbar \Omega_\lambda v_\lambda(z_1),
\label{Bog2}
\end{eqnarray}
\end{widetext}
where $\mathcal{L} =K^0_b +g_{bb} n_b +g_{bf} n_f$.
The $\hbar \Pi$ denotes the energy-representation of 
the polarization potential in Eq.~(\ref{EqB}): 
\begin{eqnarray}
     && \hbar \Pi(z_1,z_2;\Omega_\lambda)=
          g_{bf}^2 \sum_{p,h} \Phi(z_1) \Phi(z_2) 
\nonumber\\
     &&\qquad\times \left[ \frac{\psi_h^*(z_1) \psi_p(z_1) \psi_p^*(z_2) \psi_h(z_2)}{
                       \hbar \Omega_\lambda -\epsilon_p^f +\epsilon_h^f }
         \right. 
\nonumber\\
     &&\qquad\qquad 
          \left. -\frac{\psi_p^*(z_1) \psi_h(z_1) \psi_h^*(z_2) \psi_p(z_2)}{
                       \hbar \Omega_\lambda -\epsilon_h^f +\epsilon_p^f}
          \right],\quad
\end{eqnarray}
where the index $p(h)$ represents a state with energy above
(below) the fermi energy. Since we consider the stationary condensates,
the order parameter $\Phi$ is assumed to be real \cite{FR98}.

We should note that the number of eigenvalues $\Omega_\lambda$ 
obtained from Eqs.~(\ref{Bog1},\ref{Bog2}) may exceed
the dimension of the bosonic quasi-particle space
because these equations are not linear for the
eigenvalue $\Omega_\lambda$ 
(The $\hbar \Pi$ depends on the eigenvalue). 
In this case, the eigenfunctions for different eigenvalues
do not satisfy the orthogonality relations, but 
they are proved still to be a complete set.

For normalizations of the quasi-particle wave functions, 
we take
\begin{eqnarray}
     && \int\!dz_1\,\left[ |u_\lambda(z_1)|^2 -|v_\lambda(z_1)|^2 
                   \right]
\nonumber \\
     &&  -\int\!dz_1 dz_2\,\left\{ u_\lambda^*(z_1) -v_\lambda^*(z_1)
                                \right\}
\nonumber\\
     &&\quad  \times\left.\frac{d\Pi(z_1,z_2;\Omega)}{d\Omega}\right|_{\Omega_\lambda}
               \left\{ u_\lambda(z_2) -v_\lambda(z_2) 
              \right\} =1.
\end{eqnarray}

\section{\label{sec:uniform}Peierls instability of homogeneous state}

\subsection{Dispersion equation for Excitation energies}

Let's consider the case of $\omega_a=0$, 
and discuss the stability of the 1D homogeneous states in uniform mixtures, 
where the boson/fermion (1D) densities $n_{b,f}$ and the order parameter $\Phi$ 
take constant values:
\begin{equation}
     \Phi=\sqrt{\frac{N_b}{L}},  \quad
     n_{b,f} =\frac{N_{b,f}}{L},
\label{EqA}
\end{equation}
where $L$ is a quantization length for the box-potential regularization. 
In this case, the fermion single wave functions become
plane waves, $\psi_k=e^{ikz}/\sqrt{L}$, 
with the wave number $k =\pi n/L$ ($n={\rm integer}$)  
and the mean-field single-particle energy 
$\epsilon^f_k =\hbar^2 k^2 / 2 m_f +g_{bb} n_b$. 
The fermi wave-number and energy, $k_{F}$ and $\epsilon_{F}$, becomes
$k_{F}=\pi n_{f}$ and 
$\epsilon_{F} =\hbar^2 \pi^2 n_f^2 / (2 m_{f}) +g_{bb} n_{b}$.

The boson quasi-particle wave functions $u_{\lambda}$ and $v_{\lambda}$ 
also become plane waves:
$u_{\lambda,k}(z)=u_{\lambda,k}e^{ikz}$,
$v_{\lambda,k}(z)=v_{\lambda,k}e^{ikz}$.
Substituting them into Eqs.~(\ref{Bog1},\ref{Bog2}),
we obtain
\begin{eqnarray}
     (\epsilon^b_k -\mu_b)u_{\lambda,k}
          &+&\left\{ g_{bb} n_b +\hbar \Pi_k (\Omega_\lambda) \right\}
           \{ u_{\lambda,k} -v_{\lambda,k} \}
\nonumber\\ 
          &&\qquad\qquad =\hbar \Omega_\lambda u_{\lambda,k},
\end{eqnarray}
\begin{eqnarray}
     (\epsilon^b_k -\mu_b)v_{\lambda,k}
         &+&\left\{ g_{bb} n_b +\hbar \Pi_k(\Omega_\lambda) \right\}
          \{ v_{\lambda,k} -u_{\lambda,k} \}
\nonumber\\ 
     &&\qquad\qquad =-\hbar \Omega_\lambda v_{\lambda,k},
\end{eqnarray}
where $\epsilon^b_{k} =\hbar^2 k^2/2m_{b} +g_{bb} n_{b} +g_{bf} n_{f}$,
and the polarization potential $\hbar \Pi_k(\Omega_\lambda)$ for plane waves is given by
\begin{eqnarray}
     \hbar \Pi_k(\Omega_\lambda) =g^2_{bf} n_b \int^{k_F}_{-k_F} \! \frac{dq}{2\pi}
          \left[ \frac{1}{ \hbar \Omega_\lambda -\epsilon^f_{q+k} 
                +\epsilon^f_q +i \eta }\right.,     \nonumber\\
     \left.-\frac{1}{ \hbar \Omega_\lambda -\epsilon^f_{q}
                     +\epsilon^f_{q+k} +i \eta } \right].
\label{Polhomo}
\end{eqnarray}
Using the equality of the bosonic chemical potential 
and the interaction energy in the mean-field approximation, 
$\mu_b=g_{bb}n_b+g_{bf}n_f$, 
we obtain the dispersion equation for $\Omega_\lambda$:
\begin{equation}
     ( \hbar \Omega_\lambda )^2 =(e^b_k)^2 +2e^b_k 
          \{ g_{bb} n_b +\hbar \Pi_k(\Omega_\lambda) \},
\label{bosonspec}
\end{equation}
where $e^b_{k} =\hbar^2 k^2 / 2 m_{b}$.

Introducing the complex energy 
$\Omega_{\lambda} =\omega_{\lambda} -i \gamma_{\lambda}$, 
we separate the polarization potential into the real and
the imaginary parts 
\begin{equation}
     \hbar \Pi_k(\omega_\lambda,\gamma_\lambda) 
          =\hbar \Pi^{Re}_k (\omega_\lambda,\gamma_\lambda)
          +i \hbar \Pi^{Im}_k(\omega_\lambda,\gamma_\lambda).
\end{equation}
After execution of the $q$-integration in Eq.~(\ref{Polhomo}), 
they become \cite{HW74}
\begin{widetext}
\begin{eqnarray}
     \hbar \Pi^{Re}_{k}(\omega_{\lambda},\gamma_{\lambda}) &=& 
          -\frac{A_k}{2}
          \left[ \ln{ \frac{\{ (k+2k_{F}) -\frac{2m_{f} \omega_{\lambda}}{\hbar k} \}^2
                +( \frac{ 2m_{f} \gamma_{\lambda}}{\hbar k})^2 }
     { \{ (k -2 k_{F}) +\frac{2m_{f} \omega_{\lambda}}{\hbar k} \}^2
      +( \frac{2m_{f}\gamma_{\lambda}}{\hbar k})^2} } \right.\nonumber
     \left.+\ln{
          \frac{\{ (k +2 k_{F}) +\frac{2 m_{f} \omega_{\lambda}}{\hbar k} \}^2
     +( \frac{2 m_{f}\gamma_{\lambda}}{\hbar k})^2 }
       { \{ (k -2 k_{F}) -\frac{ 2m_{f} \omega_{\lambda}}{\hbar k} \}^2
        +( \frac{2m_{f} \gamma_{\lambda}}{\hbar k} )^2 }
       } \right], 
\label{cmprealpol}\\
     \hbar \Pi^{Im}_{k}( \omega_{\lambda}, \gamma_{\lambda} ) &=&
          -A_{k} \left[ \pi \theta\left(2 k_F 
                         -\left| k -\frac{2m_{f} \omega_{\lambda}}{\hbar k} \right| 
                                  \right)
                       -\pi \theta\left(2 k_F
                         -\left| k +\frac{2m_{f} \omega_{\lambda}}{\hbar k}\right|
                                  \right)
                 \right. \nonumber\\
          &&+\arctan{ \left( 
                    \frac{2 (\frac{2m_{f}}{\hbar k})^2 
                    \omega_{\lambda} \gamma_{\lambda} }{
                    ( k + 2k_{F} )^2 +( \frac{2m_{f}}{\hbar k})^2 
                                      ( \gamma_{\lambda}^2 -\omega_{\lambda}^2 ) } 
                    \right) 
                  }
   \left.-\arctan{ \left(
                   \frac{2 (\frac{2m_{f}}{\hbar k})^2
                   \omega_{\lambda} \gamma_{\lambda} }{
                   ( k -2 k_{F} )^2 +( \frac{2m_{f}}{\hbar k})^2
                                     ( \gamma_{\lambda}^2 -\omega_{\lambda}^2 ) }
                   \right)
                 }
          \right],
\label{cmpimagpol}
\end{eqnarray}
\end{widetext}
where
$A_{k} =\frac{g^2_{bf} n_b}{2\pi} \frac{m_{f}}{\hbar^2k}$,
and the principal values should be taken for the arctangent function:
$-\frac{\pi}{2} \leq \arctan{x} \leq \frac{\pi}{2}$.
Using these results, Eq.~(\ref{bosonspec}) 
can be separated into the coupled equations
\begin{eqnarray}
     \hbar^2 ( \omega_{\lambda}^2 -\gamma_\lambda^2 )
          &=& ( e^b_{k} )^2 
          +2 e^b_{k} \{ g_{bb}n_b 
                       +\hbar \Pi^{Re}_{k} \},
\label{comperg1}\\
     \hbar^2 \omega_{\lambda} \gamma_{\lambda}
          &=& e^b_{k} \cdot \hbar \Pi^{Im}_k.
\label{comperg2}
\end{eqnarray}

Now, in order to analyze the dispersion equations 
(\ref{comperg1},\ref{comperg2}) 
qualitatively, 
we discriminate two cases. 
First, when $\hbar\Pi^{Im}_k(\omega_\lambda)\neq 0$
and $\omelam, k >0$, 
Eqs.~(\ref{comperg1}) and (\ref{comperg2}) have solutions 
with complex energies ($\gamma_{\lambda}>0$), 
which correspond to the continuum p-h excitations
in the continuum. 

Second, in the region where any p-h pair excitations are prohibited
($\gamma_{\lambda}=0$),
the polarization potential becomes real:
\begin{equation}
     \hbar \Pi^{Re}_k( \omega_\lambda )
          =-A_{k} \ln{ \left| \frac{ ( k +2 k_F )^2 
                                    -( 2 m_{f} \omega_{\lambda} / \hbar k )^2 }{ 
                                     ( k -2 k_F )^2 
                                     -( 2 m_{f} \omega_{\lambda} / \hbar k )^2
                                   }
                       \right| }.
\label{rePi}
\end{equation}
In this case, introducing the dimensionless variables 
$\tilde{\omega}_\lambda =2\omelam/(v_F k_F)$ 
and $\tilde{k}=k/k_F$ 
($v_F=\hbar k_F/m_{f}$: the fermi velocity),
the dispersion relation, eq.~(\ref{comperg1}), 
becomes
\begin{equation}
     \tilde{\omega}_\lambda^2 
          =\frac{ m^2_{f} }{ m^2_{b} } \tilde{k}^4 
          +2 \tilde{k}^2 \frac{v_B^2}{v_F^2}
            \left\{ 2 -\frac{\zeta}{\tilde{k}}
                      \ln{ \left| \frac{ \tilde{k}^2 ( \tilde{k} +2 )^2
                          -\tilde{\omega}_\lambda^2 }{
                          \tilde{k}^2 ( \tilde{k} -2 )^2
                          -\tilde{\omega}_\lambda^2
                                       }
                           \right| 
                         }
            \right\},
\label{nlineq}
\end{equation}
where $v_B=\sqrt{g_{bb}n_b/m_{b}}$ and
 $\zeta$ is the dimensionless boson-fermion coupling constant defined by
\begin{equation}
     \zeta =\frac{ g_{bf}^2 }{ \pi g_{bb} \hbar v_F }.
\label{zeta}
\end{equation}
The solutions of Eq.~(\ref{nlineq}) correspond 
to the bosonic excitation energies, 
which we are interested in, 
and $v_B$ is just the Bogoliubov-phonon sound velocity. 
It should be noted that the scaled equation (\ref{nlineq}),
depends on three dimensionless parameters,
$m_{f}/m_{b}$, $v_B/v_F$, and $\zeta$.

\subsection{Excitation Spectra and Peierls Instabilities}

Let's study the bosonic excitation spectra 
on the homogeneous ground state (\ref{EqA}) 
with solving the dispersion relations 
(\ref{comperg1},\ref{comperg2}). 
To understand general features, 
we show numerical results for two typical cases, 
(a) $v_B < v_F$  and (b) $v_B > v_F$, in Figure.~2:
(a) $(v_B/v_F)^2=0.67$, $\zeta=0.99$ and 
(b) $(v_B/v_F)^2=1.50$, $\zeta=0.80$. 
The same values have been taken for boson and fermion masses,
$m_b =m_f$, for all cases. 
The shaded areas in Figure~2 correspond 
to the continuum spectra corresponding to 
the fermionic p-h excitation states, 
and the solid lines are to the isolated modes,
corresponding to the collective excitations.
It should be noted that two collective modes 
(low- and high-lying)
exist with the same wave number $k$ 
(except narrow regions around $2 k_F$);
they can be interpreted to be coherent superposition 
of the Bogoliubov-phonon mode from bosonic atoms
and the fermion p-h collective mode. 
For comparison, 
the Bogoliubov-phonon spectra 
in boson-fermion noninteracting systems ($g_{bf}=0$) 
have been plotted in Figure~2, 
and we can find that it runs between two collective spectra 
in the interacting ones.
\begin{figure}
\begin{center}
\includegraphics[width=8cm,height=11cm]{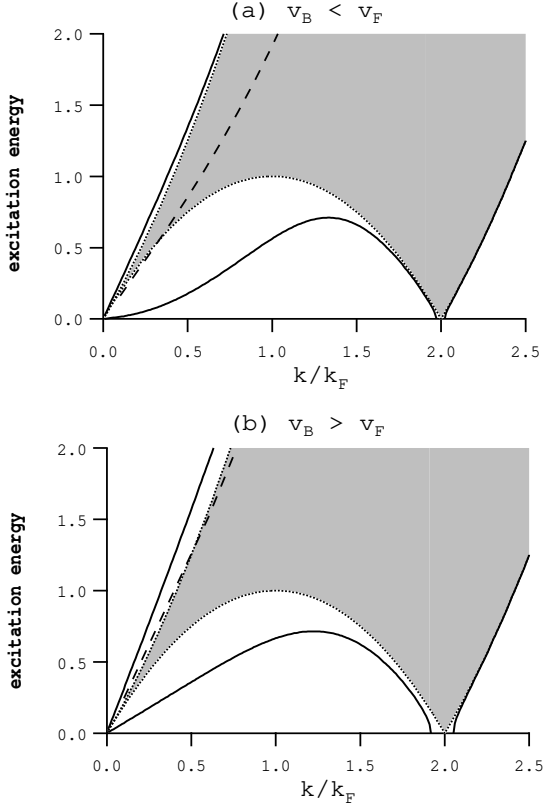}
\vspace{.3 cm}
\caption{\label{fig:2} Excitation spectrum (in units of $\hbar v_F k_F$)
of bose-fermi mixtures.
Solid lines: collective modes of p-h pairs and bosonic modes,
dashed line: free boson,
shaded area: continuum p-h pair excitation spectra.}
\label{spectrum}
\end{center}
\end{figure}

Let's discuss the specific features of the collective modes, 
especially in small $k$ and $k \sim 2 k_F$ regions.

In small wave-number region, $k < k_F |1-(v_B/v_F)^2|$, 
the Bogoliubov-phonon modes (dashed line) 
in the boson-fermion noninteracting limit are located 
below the fermion p-h excitation modes in energy 
and absorbed into the fermion p-h continuum
when $v_B < v_F$ (case (a)). 
In case that  $v_B > v_F$ (case (b)), 
the corresponding modes are above the continuum.
It suggests that the low-lying collective mode in (a) 
is mainly a Bogoliubov-phonon mode from bosonic atoms
but the corresponding mode in (b) is mainly 
the fermion p-h one, 
and the high-lying modes have opposite characters. 
In Figure~2, the bosonic excitation modes are found 
to be influenced by the interaction 
with the fermion p-h excitation modes, 
and pushed downward/upward in case (a)/(b). 

Around $k \simeq 2 k_{F}$, 
the low-lying collective modes become very soft and, 
between two critical wave-numbers $k_{-} < k < k_{+}$, 
they are found to be zero-modes in both cases.
To calculate the energy of these modes analytically, 
we expand Eq.~(\ref{comperg2}) to the leading order 
of $|q|/k_F$ ($q =k -2 k_{F}$):
\begin{eqnarray}
     \hbar^2\omega_{\lambda} \gamma_{\lambda}
          &=& A_{2k_{F}} e^b_{k_{F}}
           \left[ \pi \theta( \omega_{\lambda} -v_{F} |q| ) \right.
\nonumber\\
          &&\left. -\arctan\left( \frac{2\omega_\lambda\gamma_\lambda}{
                                 v_F^2 q^2 -\omega_\lambda^2 
                                 +\gamma_\lambda^2 
                                }
                   \right) \right].
\end{eqnarray}
When $\omega_{\lambda} < v_{F}|q|$,
the above equation reduces to $\omega_{\lambda} \gamma_{\lambda} =0$.
In the case of $\omelam=\gamlam= 0$, 
two critical wave-numbers are obtained from Eq.~(\ref{comperg1}): 
\begin{equation}
     \frac{k_{\pm}}{k_{F}} =2 \pm 4
          \exp\left[ -\frac{2}{\zeta}
                   \left\{1 +\left( \frac{m_{f}v_{F}}{m_{b}v_{B}} \right)^2
                  \right\}
             \right],
\label{eqA}
\end{equation}
which gives ($k_{-}=1.97k_F$, $k_{+}=2.02k_F$) for (a) in Figure~2, 
and ($k_{-}=1.92k_F$, $k_{+}=2.05k_F$) for (b) in the same figure.

Analyzing the eigenvalues of Eq.~(\ref{comperg1}) more detail, 
we find that they become purely imaginary
($\omelam=0$, $\gamlam\neq 0$)
in the region between the two critical wave-numbers, $k_{-} < k < k_{+}$; 
it is just the ``Peierls instability'' in Q1D system,  
which suggests that the homogeneous state is unstable 
against the fluctuations with the wave-number around $2k_F$.

Finally, we should comment on a different kind of instability 
that may occur in the small wave-number fluctuation \cite{Das}. 
When we take the $k \to 0$ limit with keeping
$\omega_\lambda / k ={\rm constant}$, 
the real part of the polarization potential becomes
\begin{equation}
     \hbar\Pi^{Re}_k(\omega_\lambda) 
          =\frac{ ( g_{bf} )^2 n_b }{ \pi \hbar v_F }
           \frac{ ( v_F k )^2 }{ \omega_\lambda^2 -( v_F k )^2 }.
\label{k0Pilimit}
\end{equation}
and the imaginary part $\Pi^{Im}_k$ vanishes
except at $v_B =v_F$.
Substituting Eq.~(\ref{k0Pilimit}) into Eq.~(\ref{comperg1}),
we obtain two branches of excitations: 
\begin{equation}
     \hbar \omega_{\pm} =\hbar k 
          \left[ \frac{ v_F^2 +v_B^2 }{2}
                \pm\frac{1}{2} \sqrt{ ( v_F^2 -v_B^2 )^2
                +4 \zeta v_F^2 v_B^2}
         \right]^{1/2}.
\end{equation}
In the strong boson-fermion interacting case ($\zeta>1$),
the energy of the low-lying mode becomes purely imaginary, 
and it suggests an instability of the system.
We find that the stability condition for the $k \simeq 0$ mode: 
\begin{equation}
     \frac{\pi^2 \hbar^2}{m_{f}} n_f \ge \frac{ ( g_{bf} )^2 }{ g_{bb} },
\label{PSC}
\end{equation}
where we have used the relation $n_f =k_F / \pi$. 
This $k \simeq 0$ instability is considered to cause 
the phase-separated states more stable. 
In the present paper, 
we do not discuss this instability 
and concentrate only on the density-wave states 
caused by the Peierls instability, 
which is discussed in the next section.  
It should be a very interesting problem 
to study the cooperation/competition between these states. 

\section{\label{sec:periodic} DENSITY WAVES IN BOSON-FERMION GROUND STATE}

The results obtained in the previous section shows 
that the homogeneous state is not
the true ground state of the system 
and a lower energy can be obtained for the state
with a spatially-periodic condensate
characterized by the $2k_F$-periodic order parameter: 
$\Phi(z) =b_u +b_p \cos{(2k_Fz)}$.
The constant bosonic density is given by $n_b=b_u^2+b_p^2/2$.
In the case of weakly-periodic variation $b_p\ll b_u$,
up to second order in $b_p$, the mean-field Hamiltonian
for the bose-fermi system becomes
\begin{eqnarray}
     \frac{\hat{H}}{L} &=&\frac{g_{bb}}{2} (b_u^4+b_u^2b_p^2)
                                 +\epsilon^b_p b_p^2
                                 +\sum_{k} \epsilon^f_k 
                                  \hat{c}_k^\dagger \hat{c}_k
\nonumber\\
                                 &+& g_{bf} b_u b_p
                                     \sum_k ( \hat{c}^\dagger_{k -2 k_F}
                                              \hat{c}_k
                                             +\hat{c}^\dagger_{k+2k_F}
                                              \hat{c}_k),
\label{perihami1}
\end{eqnarray}
where $\hat{c}_k$ and $\hat{c}_k^\dagger$ are 
the annihilation and creation operators 
for fermions with the wave-number $k$ 
and the energy
$\epsilon^f_k =\hbar^2 k^2 /2 m_f +g_{bf} (b_u^2+b_p^2/2)$.
The $\epsilon^b_p$ in Eq.~(\ref{perihami1}) 
is the energy generated from the periodic condensate: 
$\epsilon^b_p =e^b_{2k_F} /2+g_{bb} b_u^2$.
The last term in Eq.~(\ref{perihami1}) 
is for the processes 
that the fermion with a wave-number $k$ 
feeling periodic- and uniform-bosonic condensates
scatters into the sates with $k\pm 2k_F$.

In order to obtain a single-particle energy, 
we calculate the fermion Green's function:
\begin{equation}
     i F_{k,k'}(t -t') 
       \equiv 
          \bra T[ \hat{c}_k(t) \hat{c}^\dagger_{k'}(t') ] \ket
       =\int \frac{d\omega}{2\pi} e^{-i \omega (t -t')} 
                                  F_{k,k'}(\omega),
\end{equation}
where 
$\hat{c}_k(t) =e^{i \hat{H} t /\hbar } \hat{c}_k e^{ -i \hat{H} t / \hbar}$.
For a fixed wave-number $k$, 
the Green's function $F_{k,k'}(\omega)$ has 
non-vanishing off-diagonal matrix elements at $F_{k,k\pm 2k_F}$ 
in the Hamiltonian (\ref{perihami1}). 
In the limit $k \to +k_F$, 
fermion states with wave-numbers $k$ and $k -2k_F$ are almost degenerate
in energy, 
but that with $k+2k_F$ is separated.
Consequently, 
in case of $k_F\ge k >0$, 
the Dyson equation for the Green's function can be approximated by
\begin{equation}
     \left( \begin{array}{cc}
               \hbar \omega -\epsilon^f_k & -\Delta \\
                                  -\Delta & \hbar \omega -\epsilon^f_{k -2 k_F}
            \end{array}
     \right)
     \left( \begin{array}{c} F_{k,k} \\ F_{k,k -2 k_F} \end{array} \right)
     =\left( \begin{array}{c} \hbar \\ 0 \end{array} \right),
\label{dysonmtx}
\end{equation}
where $F_{k,k}$ is a diagonal part of the Green's function, 
and $\Delta =g_{bf} b_u b_p$.
They can be solved easily:
\begin{eqnarray}
     F_{k,k} &=&\frac{ U_k^2 }{ \omega -E^{-}_{k} / \hbar -i \eta}
             +\frac{ V_{k}^2 }{ \omega -E^{+}_k / \hbar +i \eta}, 
\label{offdiag2}\\
     F_{k,k -2 k_F} &=&\frac{ -U_k V_k }{ \omega -E^{-}_{k} / \hbar -i\eta}
                    +\frac{ U_k V_{k} }{ \omega -E^{+}_k / \hbar +i \eta}, 
\label{offdiag1}
\end{eqnarray}
where
\begin{eqnarray}
     U_k &=&\sqrt{ \frac{1}{2} } 
          \left\{ 1 -\frac{ \xi_k }{ \sqrt{\xi^2_k +\Delta^2} }
          \right\}^{1/2}, \label{EqU}\\
     V_k &=&\sqrt{ \frac{1}{2} }
          \left\{ 1 +\frac{ \xi_k }{ \sqrt{\xi^2_k +\Delta^2} }
         \right\}^{1/2}, \label{EqV}
\end{eqnarray}
with $\xi_k =\hbar v_F(k -k_F)$.
The single-particle energy with wave-number $k$ 
is obtained from the pole of $F_{k,k}(\omega)$: 
\begin{equation}
    E^{\pm}_k =\epsilon^f_k -\xi_k \pm \sqrt{\xi_k^2 +\Delta^2},
\label{EqF}
\end{equation}
from which we can understand 
that the $2 |\Delta|$ is the energy gap at the fermi surface.
It is clear that the periodic condensate has produced an energy gap $2 |\Delta|$ 
near the fermi surface.

To evaluate the value of $\Delta$, 
we calculate the total energy of the system 
in the present approximation.
The fermionic contributions to the energy 
is obtained by summing-up the $E_k^-$ in Eq.~(\ref{EqF}) 
up to the fermi wave-number $k_F$: 
\begin{eqnarray}
     E_f &=&\int dk E^{-}_k \theta(k_F-|k|)
          = \left( \frac{ e_F }{ 3 } +g_{bf} n_b
            \right) n_f 
\nonumber\\
         &+& n_f \Bigg[ e_F -\sqrt{e_F^2 +\Delta^2} 
\nonumber\\
         &&\qquad\quad -\frac{ \Delta^2 }{ e_F } 
                     \ln{ \left( \frac{e_F +\sqrt{e_F^2 +\Delta^2}}{|\Delta|}
                         \right)
                        }
              \Bigg],
\end{eqnarray}
where $e_F =\hbar v_F k_F/2$.
For simplicity, we consider the weak coupling limit, 
$\Delta \ll e_F$, then the total energy density of the system becomes
\begin{eqnarray}
     && E_{tot}(|\Delta|) -E_{tot}(0) =\frac{n_f \Delta^2}{4 e_F} 
\nonumber\\ 
     && \times \left[\frac{2}{\zeta}
              \left\{ 1 +\left( \frac{m_f v_F}{m_b v_B} \right)^2
             \right\}
              -\frac{1}{2} \left\{ 1 +2\ln\frac{4e_F}{|\Delta|}
                          \right\}
        \right].
\label{EqG}
\end{eqnarray}
Minimizing Eq.~(\ref{EqG}) with respect to $|\Delta|$, 
we obtain the stationary value of the gap parameter
\begin{equation}
     \frac{|\Delta|}{e_F} 
          =4 \exp{ \left[ -\frac{2}{\zeta} 
                        \left\{ 1 +\left( \frac{m_f v_F}{m_b v_B} \right)^2
                       \right\}
                  \right]
                 }.
\label{GAPE}
\end{equation}
Using the above result, 
the energy difference between the ground 
and the uniform ($\Delta =0$) states is
\begin{equation}
     E_{tot}(|\Delta|) -E_{tot}(0) 
          =-\frac{\Delta^2}{4\pi \hbar v_F},
\end{equation}
which shows that the state with the ``periodic condensate'' ($b_p \neq 0$) 
has lower energy than the uniform one. 

\begin{figure}
\begin{center}
\includegraphics[width=8cm,height=5cm]{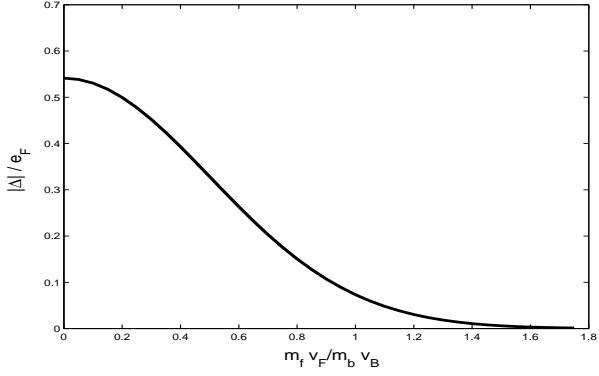}
\vspace{.3 cm}
\caption{\label{fig:3} Gap energy as function of
$m_{f}v_{F}/m_{b}v_{B}$ (in units of $e_F$)
for $\zeta=0.99$.}
\label{gap}
\end{center}
\end{figure}

Next, we discuss the fermionic density, 
which is obtained from the Green's function: 
\begin{eqnarray}
     n_f(z) &=&\lim_{\eta\to 0_+} 
             \sum_k \int \frac{d\omega}{2\pi i}
                    e^{i\eta\omega}
             \Big[ F_{k,k} +F_{k,k -2 k_F} e^{i2k_Fz} 
\nonumber\\
            &&\quad +F_{k -2 k_F,k} e^{-i 2 k_F z}
                   +F_{k -2 k_F,k -2 k_F}
             \Big]. 
\end{eqnarray}
Using Eqs.~(\ref{offdiag2},\ref{offdiag1},\ref{EqU},\ref{EqV}), 
it becomes
\begin{eqnarray}
\label{denwave}
     n_f(z) &=&\sum_k[ U_k^2 +V_k^2 -2 U_k V_k \cos{(2 k_F z)}] 
\nonumber\\
            &=&n_f \left[ 1 -\frac{\Delta}{e_F \zeta}
                             \cos{(2k_Fz)} 
                   \right].
\label{EqNF}
\end{eqnarray}
which shows that, for finite $\Delta$,  
the fermion state is in the density-wave one
where the fermionic density has a periodicity $2k_F$.
Corresponding to Eq.~(\ref{EqNF}), 
the direct calculation shows 
that the bosonic density $n_b(z)$ has the similar density-wave structure: 
\begin{equation}
     n_b(z) =n_b \left[ 1 +\frac{|\Delta|}{|g_{bf}| n_b} \cos(2k_Fz)
                \right].
\label{EqNB}
\end{equation}
It should be noted that, 
different from $n_f(z)$, 
the periodic part in $n_b(z)$ is proportional to the absolute value 
of the gap parameter $|\Delta|$.  
That means that the boson and fermion density-waves 
are out of phase for the boson-fermion repulsive system, 
but in phase for the attractive one. 

As seen in Eq.~(\ref{GAPE}), 
the gap energy $\Delta$ essentially depends on two combinations of parameters, 
$m_f v_F / m_ bv_B$ and the dimensionless coupling constant $\zeta$. 
In Figure.~\ref{gap}, 
we just plotted the gap energy 
as a function of $m_f v_F /m_b v_B$ for $\zeta =0.99$, 
from which we can read off 
that the boson-fermion density-wave state can be observed 
in increasing the 1D density of condensate
and/or using heavier bosons than fermions.

\section{\label{sec:finite}PEIERLS INSTABILITY OF SYSTEM 
IN HARMONIC OSCILLATOR POTENTIAL}

In this section, 
we briefly discuss the non-uniform boson-fermion mixtures 
confined in a HO potential
with the finite axial dimension $(\omega_a\ne 0)$,
and the condition for its Peierls instability. 

Using the eigenenergy 
$\epsilon^{ho}_{n} =\hbar \omega_{a}(n +\frac{1}{2} )$
($n =0,1,2, \cdots$) and the wave function $\phi^{ho}_{n}$
for the 1D HO potential,
the quasiparticle amplitudes
$u_{\lambda}$ and $v_{\lambda}$ are expanded by
\begin{equation}
     u_{\lambda}(z) =\sum_{n} u^\lambda_{n} \phi^{ho}_{n}(z),  \quad
     v_{\lambda}(z) =\sum_{n} v^\lambda_{n} \phi^{ho}_{n}(z).
\end{equation}
Substituting them into Eqs.~(\ref{Bog1},\ref{Bog2}),
we can obtain the eigenequations for the excitation energies
$\hbar\omega_{\lambda}$:
\begin{eqnarray}
     \sum_{n} \left[ ( \epsilon^{ho}_{n} -\hbar\omega_\lambda ) 
                     \delta_{m,n}u^\lambda_n
                    +\hbar \Pi_{mn}( \omega_\lambda )
                           ( u^\lambda_n -v^\lambda_n )
              \right] =0, 
\label{RPA1}\\
     \sum_{n} \left[ ( \epsilon^{ho}_{n} +\hbar\omega_\lambda )
                     \delta_{m,n}v^\lambda_n
                    +\hbar \Pi_{mn}( \omega_\lambda )
                           ( v^\lambda_n- u^\lambda_n )
            \right] =0.
\label{RPA2}
\end{eqnarray}
where, to concentrate on the role of the polarization potential effects 
in finite systems, 
we have neglected the Hartree potential.
The polarization potential $\hbar\Pi_{mn}$ in Eqs.~(\ref{RPA1},\ref{RPA2}),
is given by
\begin{eqnarray}
     && \hbar \Pi_{mn}(\omega_\lambda) =(g_{bf})^2 N_b
\nonumber\\
     &&\times \sum_{ph} \left[ \frac{ {\langle m h|p 0\rangle} 
                                  {\langle 0 p|h n \rangle} }{
                                  \hbar\omega_\lambda
                                 -\epsilon^{ho}_{p}
                                 +\epsilon^{ho}_{h}
                                }
                    -\frac{ {\langle m p|h 0\rangle} 
                             {\langle 0 h|p n\rangle} }{
                             \hbar\omega_\lambda
                            -\epsilon^{ho}_{h}
                            +\epsilon^{ho}_{p}
                           }
             \right].
\label{PP}
\end{eqnarray}
The matrix elements in Eq.~(\ref{PP}) are defined by
\begin{equation}
     {\langle mh |p0 \rangle} = \int\! {\mathrm d}z \phi^{ho}_{m} \phi^{ho}_{h}
                                                    \phi^{ho}_{p} \varphi_{0},
\label{EqME}
\end{equation}
where the single-particle wave functions in Eq.~(\ref{EqME}) have been
replaced by the HO eigenfunctions $\phi^{ho}_{n}$. 
It should be noted that the matrix element
Eq.~(\ref{EqME}) has a transition probability 
only in the spatial region
where condensate exists.
It is important at this point to contrast finite system to
homogeneous system in which condensate is present all
over the range. 

We restrict the present discussion into the parameter region:
$\hbar\omega_{r} >\mu_{f} >\mu_{b} >\hbar\omega_{a}$.
Under these conditions, we can take several approximations.
Because of $\mu_{b} > \hbar \omega_{a}$, 
the system is in the Thomas-Fermi (TF) regime, 
so that we can take the TF approximation 
for the order-parameter:
\begin{equation}
     \Phi =\sqrt{N_b} \varphi_{0} 
          \propto \sqrt{ 1 -\frac{z^2}{z^2_{TF}}}
                  \theta(z_{TF}-|z|).
\label{EqTF}
\end{equation} 
The $z_{TF}$ in Eq.~(\ref{EqTF}) is a cut-off length 
in the TF approximation:
\begin{equation}
     z_{TF}  =\sqrt{2} a_{ho} \left(\frac{3N_b \beta}{4\sqrt{2}}\right)^{1/3},
\end{equation} 
where 
$\beta =mg_{bb} a_{ho} /\hbar^2$
with
$a_{ho}=\sqrt{\hbar / m \omega_{a}}$ 
(the HO length in the axial direction).

As we have seen in chapter III, 
Peierls instability comes from the coupling 
between the boson mode with wave-number near $2k_F$ 
and the fermion p-h excitations near $k_F$.
Although the wave-number is not strictly conserved in a HO
potential, the instability mode would have 'wavenumber':
$k_n= \sqrt{2n}/a_{ho}\sim 2k_F$.
Since the higher nodal modes $n\sim N_f$ have a broader
spatial extension ($L_n=\sqrt{2n}a_{ho}$) than a condensate one,
the asymptotic expansions can be used for the $\phi^{ho}_{n}$
in the middle of the trap $|z| < z_{TF} < L_n$
\cite{GW00}:
\begin{equation}
     \phi^{ho}_n(z) \propto
                    \cos{ \left( k_n z -\frac{n\pi}{2} 
                          \right)
                        }.
\end{equation}
For a bosonic eigenstate with higher nodal mode,
the negative energy amplitude $v^\lambda_{n}=0$
of Eq.~(\ref{RPA1}) can be neglected.

Using these approximations, 
the matrix elements like $\langle 0 p | h n \rangle$ become
linear combinations of the form $J_1(z_{TF}K)/z_{TF}K$,
where $J_1(x)$ is the 1st-order Bessel function 
and the $K$ is a wave-number difference
between the initial and the finial states: $K=k_{n}+k_{h}-k_{p}$.
In the case of $|x| \gg 1$, 
using the asymptotic expansion of the Bessel function, $J_1(x)/x \sim |x|^{-3/2}$,
the terms of $|x|\leq {\mathcal O}(1)$
can contribute to the matrix elements.
Under the restriction of wave-numbers $k_h \leq k_F$ and $k_p> k_F$
with $k_n\sim 2k_F$ ($k_F=\sqrt{2N_f}/a_{ho}$), 
the main contributions of matrix element are 
from the scattering processes: 
$k_n \to k_p +k_h$ and $k_n +k_h\to k_p$.
To make an estimate,
we replace a product of Bessel functions as
\begin{equation}
     \frac{J_{1}(x_{1})}{x_{1}} \frac{J_{1}(x_{2})}{x_{2}}
          \simeq \frac{1}{4} \theta(1 -|x_{1}|) \theta(1 -|x_{1} -x_2|),
\end{equation}
and take the continuum-limit for the wave-number sums: 
$\sum_{p,h} \to \int\!{\mathrm d}p \int\!{\mathrm d}h$.

Finally, 
the eigenenergy for a collective state 
with the wave number $k_{n}=2k_{F}$ 
can be obtained 
%
%
\begin{equation}
     \hbar \omega_{\lambda} =4 \epsilon_F
                            -\frac{g_{bf}^2 n_b(0)}{2 \pi \hbar v_F}
                             \ln{|4 k_{F} z_{TF}|}.
\label{EE}
\end{equation}

Contrary to the uniform system, 
the static polarization potential, 
the second term of right hand side of Eq.(\ref{EE}), 
is not divergent at $2k_F$. 
This is because the scattering of trapped atoms
without wave-number conservation smears a singularity 
due to a sharp fermi sea. 
The Peierls instability occurs 
where the polarization energy is overcome by the kinetic energy $4\epsilon_F$:
\begin{equation}
     1 < \frac{\zeta}{4} \frac{v_B(0)^2}{v_F^2} \ln{|4 k_Fz_{TF}|}
\label{PIC}
\end{equation}
where $v_{B}(0) =\sqrt{g_{bb} n_{b}(0) /m}$.
This is the Peierls instability condition where boson-fermion
density wave may occur.

We roughly estimate the parameters 
according to experimental conditions.
When $m_b / m_F \sim 1$, 
we can express the velocity ratio
and the dimensionless coupling constant by
\begin{eqnarray}
     \frac{v^2_B}{v^2_F} &=&\frac{1}{4N_f}
          \left( 3 N_b \frac{\omega_r}{\omega_a}
                       \frac{a_{bb}}{a_{ho}}
          \right)^{\frac{2}{3}}, \\
     \zeta &=&\sqrt{\frac{2}{\pi^2N_f}}
           \frac{\omega_r}{\omega_a}
           \frac{a_{bf}^2}{a_{bb} a_{ho}}.
\end{eqnarray}
As a possible candidate for the Peierls instability, 
we take the rubidium isotope system: $^{87}$Rb$-^{84}$Rb mixtures.
Taking the scattering lengths in \cite{BB99},
$a_{bb}=5.3\,{\rm nm}$ and $a_{bf}=29.1\,{\rm nm}$ 
and
the typical trapping frequencies, 
$\omega_a=2 \pi \times 10\,{\rm Hz}$ and 
$\omega_r=2 \pi \times 15\,{\rm kHz}$, 
Eq.~(\ref{PIC}) gives $N_f=10^3$ and $N_b=2\times 10^4$ 
for the realization of the Peierls instability.

\section{\label{sec:discus}Summary}

In the present paper, 
we studied the occurrence of Peierls instability in Q1D bose-fermi mixtures 
at zero temperature.
We analyzed the bosonic collective-excitation spectra
in random phase approximation.
It shows that the mixtures of uniform BEC and a fermi gas are unstable 
against a spontaneous formation of a collective mode of wave-number $2k_F$; 
this type of instability is known as Peierls instability.
This result suggests
that the ground state of bosons
is a periodic condensate with the period $\pi/k_F$.
In the variational method, 
the boson-fermion density wave state 
have been shown to have a lower energy than the uniform state.
We also expanded our analysis for systems in an axial harmonic
oscillator potential and derived Peierls instability condition.

It is well known that density waves in Q1D conductors
lead to generations of a Lee-Rice-Anderson \cite{LRA73}
and a phase soliton \cite{1DCOND} modes. 
Thus, studies of the dynamical properties of boson-fermion density waves
should be interesting problems in future.

The authors are supported by Grant-in-Aids for Scientific Research 
by JSPAS.
One of the authors (Miyakawa) was also supported by JSPS fellowship of Japan.

\end{document}